# ANALYSIS OF EVOKED EMG USING WAVELET TRANSFORMATION

**Zaid Bin Mahbub[1], J H Karami[2] , K Siddique-e Rabbani [3]**

[1] Department of Arts & Sciences, Ahsanullah University of Science & Technology, Dhaka, Bangladesh
[2] Department of Statistics, University of Dhaka, Dhaka, Bangladesh
[3] Department of Biomedical Physics & Technology, University of Dhaka, Dhaka, Bangladesh

Email: zaidbin@gmail.com, mjhkarami2000@yahoo.com, rabbani@univdhaka.edu

**ABSTRACT:** Evoked EMG M-responses obtained from the thenar muscle in the palm by electrical stimulation of the median nerve demonstrate a well-established smooth bipolar shape for normal healthy subjects while kinks are observed in certain neurological disorders, particularly in cervical spondylotic neuropathy. A first differentiation failed to identify these kinks because of comparable values obtained for normally rising and falling segments of the smooth regions, and due to noise. In this study, the usefulness of the wavelet transform (WT), that provides localized measures of non-stationary signals is investigated. The Haar WT was used to analyze a total of 36 M-responses recorded from the median nerves of 6 normal subjects (having smooth shape) and 12 subjects with assumed neurological disorders (having kinks), for two points of stimulation on the same nerve. Features in the time-scale representation of the M-responses were studied using WT to distinguish smooth M-responses from ones with kinks. Variations in the coefficient line of the WT were also studied to allow visualization of WT at different scales (inverse of frequency). The high and low frequency regions in the WT came out distinctively which helped identifications of kinks even of very subtle ones in the M-responses which were difficult to obtain using the differentiated signal. In conclusion, the wavelet analysis may be a technique of choice in identifying kinks in M-responses in relation to time, thus enhancing the accuracy of neurological diagnosis.

## 1. INTRODUCTION

In motor nerve conduction measurements a nerve trunk supplying a group of muscles is electrically stimulated while the resulting 'evoked EMG'or the compound muscle action potential (CMAP, also called the M-response) is recorded using surface electrodes on the muscle. Conventionally a value of Motor Nerve Conduction Velocity (MNCV) is obtained from the difference of the onset latencies of direct evoked EMG by stimulating the nerve at two points on the nerve at two different distances from the recording point. It is common to observe variations in the evoked EMG waveforms for patients with neuropathy from a typical normal shape obtained from normal healthy subjects, which is expected from the genesis of CMAPS due to a large number of individual nerve fibres, whose conduction velocity profile may change with disease [1]. Therefore, diagnostic information should be hidden in the evoked EMG signal, which would be interesting to isolate and classify according to the assessment of neurological disorders. Much has been done in the analysis of 'voluntary EMG' [2-7] in order to extract diagnostic information of neurological disorders and some are used in standard clinical procedures. Furthermore, evoked nerve action potentials obtained directly from the nerves were analysed in order to obtain estimates of distribution of conduction velocity (DCV) of nerve fibres [1,8-10]. Correlation of diseases or disorders with latency values of evoked EMG responses were also studied [11-13], but none other than the group at Dhaka University analysed 'evoked EMG' waveforms [14-18] which are distinct from 'voluntary EMG' mentioned above to seek a relationship to DCV. Evoked EMG includes some uncertainties due to delays at the neuromuscular junction [19], and possibly due to this reason no other group has attempted analysing these outputs from the human body. However, a long clinical experience of the group at Dhaka University suggested that the evoked

EMG waveform or the M-response has shapes which are almost the same for subjects with normal healthy nerves, and deviations from this shape would indicate neuropathy, in general.

They argued, differences in the evoked EMG waveforms obtained from the same recording electrodes for two or more points of stimulation on the supplying nerve would be related only to the DCV of the nerve trunk. Therefore there could be a possibility of extracting DCV information, at least grossly, from the evoked EMG waveforms obtained from several points of stimulation on the same nerve. Initially this group performed synthesis of CMAPs or evoked EMG M-responses through numerical simulation based on DCV, as a Forward problem. Then they used this information to extract qualitative information on DCV from the CMAPs with various assumed disorders [15-17]. One of group's significant findings was the reproducibility of the distribution of F- latencies (DFL) which was argued to be the mirror image of DCV, at least of motor nerve fibres contributing to the F-responses [19-23]. F-responses are delayed evoked EMG potentials which are known to occur because of the travel of evoked nerve action potentials to the spinal cord and random backfiring in a few percent of the cell bodies.

For a normal healthy subject the evoked EMG signal from abductor policis brevis muscle of the palm, supplied by the median nerve, has a smooth bipolar shape [24]. Based on the above mentioned simulation work the Dhaka group has shown that kinks in the evoked EMG relates to a DCV having a double peak, implying the loss of fibres in the mid-velocity range. Through later work on DFL, the group has also found evidence to relate double peaks of DFL to kinks in evoked EMG potentials, which in turn could be related to Cervical Spondylosis, though further work is necessary to confirm and specify this claim [20-23].

Moreover the group has also analyzed the evoked EMG M-responses in frequency domain using FFT. A number of parameters such as peak amplitude, peak frequency, frequency bandwidths, and areas in specified frequency segments, etc. proved their usefulness for finding the kinks objectively [18]. Since the Fourier transform has no time localization, information about the occurrence of kinks in time was absent. Using first order differentiation of evoked EMG responses it should be possible theoretically, to find the kinks as discontinuities or abrupt changes in time domain. However, slopes in normally rising and falling segments of the smooth regions are also comparable making it difficult to identify the kinks. Besides, noise associated with the signal is accentuated through differentiation, which also masks the kinks.

The objective of this paper is to investigate the kinks in evoked EMG signal in time in the presence of noise, digital or otherwise, and considered using wavelet transform (WT) techniques. WT allows precise localization of different frequency components in time, thus can possibly provide analysis of the evoked EMG responses in a more comprehensible approach. In WT methods, one needs to choose a wavelet, which matches the analysed signal well. This work considers using the Haar continuous wavelet transform (CWT) - which is perhaps one of the simplest. As shapes of the Haar wavelets have both positive and negative parts as obtained for the evoked EMG's we will use it primarily in outlining the applications of wavelet analysis in M-responses.

In order to analyze the signals, real life evoked EMG signals collected earlier that represented both normal subjects (having smooth bipolar shape) and subjects with neurological disorders (having kinks) were used. The aim was to study whether CWT displays of the evoked EMGs show up the kinks distinctively. Variations in the coefficient line of the WT were also studied to allow visualization of WT at different scales (inverse of frequency). In order to perform the above tasks, Digital Signal Processing (DSP) techniques were used and was implemented using MATLAB, a well-known mathematical software package [25].

## 2. METHODS

### 2.1 Wavelet Transform

Wavelet transforms are similar to Fourier transforms (FT) in that both seek a better representation of the original data in a transformed domain. FT works well for periodic and stationary signals where time information is not important. For non-stationary signals FT is inappropriate because it cannot express how the spectral contents of a signal vary with time. The wavelet transform (WT) overcomes these deficiencies as it has ability to elucidate simultaneously both spectral and temporal information within the signal. Wavelet transforms are broadly divided into two classes: continuous and discrete.

Let $\psi \in L^2(\Re)$ be a fixed function (analyzing wavelet or mother wavelet). The corresponding family $\{\psi_{(b,a)}; a,b \in \Re, a \neq 0; a$ is the scaling parameter and b is the translation parameter$\}$ of wavelets is the family of shifted and scaled copies of ψ defined as follows,

$$\psi_{(b,a)}(t) = \frac{1}{\sqrt{a}} \psi(\frac{t-b}{a}), t \in \Re \qquad (1)$$

Satisfying the conditions,

(i) $\int_{-\infty}^{\infty} \psi(t)dt = 0$      (Having zero mean)

(ii) $\int_{-\infty}^{\infty} \psi^2(t)dt = 1$      (Square norm one), &

(iii) $\int_{-\infty}^{\infty} t^n \psi(t)dt = 0$, n = 0,1,...,N (having vanishing moments)

As *a* varies the wavelet *ψ(b,a)* varies in width and by varying *b*, the mother wavelet is displaced in time.

The continuous wavelet transform (CWT) of a finite energy signal *f* is defined by the integral,

$$C_{a,b} = \langle f, \psi_{(b,a)} \rangle = \frac{1}{\sqrt{a}} \int f(t) \psi^*(\frac{t-b}{a}) dt \qquad (2)$$

Where * denotes complex conjugation. The wavelet coefficient, $C_{a,b}$ contains information concerning the signal *f* at the scale *a* around the point *b*, indicating how precisely the wavelet function locally fits the signal at every scale *a*.

The wavelet coefficients are real numbers usually shown by the intensity of a chosen color, against a two dimensional plane. In the CWT the wavelet coefficients are evaluated for infinitesimally small shifts of translation as well as scale factors. That is, the color intensity of each pixel in the wavelet transform plane is separately evaluated, and the resulting pattern contains information about the size and location of the 'event' occurring in the time domain.

Some examples of established wavelet classes are the Daubechies wavelets, Lemarie wavelets, Haar wavelets, Gabor wavelets and spline wavelets [26-28]. However, we will go with the Haar wavelets in the present work.

The Haar wavelets with orthogonal bases, is the earliest known example of a wavelet transform, and perhaps one of the simplest. The Haar function is defined as,

$$\psi(x) = \begin{cases} 1 & 0 \leq x < 1/2 \\ -1 & 1/2 \leq x < 1 \\ 0 & 0 \text{ otherwise} \end{cases} \quad (3)$$

## 2.2 Application to evoked EMG

Evoked EMG signals of Median nerves of many subjects irrespective of neurological history were used in this study. The signals were obtained using the computerised EMG equipment designed and fabricated by the Bio-Medical Physics Group, University of Dhaka, which has been providing routine clinical services for nerve conduction investigation in Bangladesh since 1988 [29].

As mentioned before, for healthy normal Median nerves the evoked EMG waveform, obtained from the Thenar muscle of the palm, has a smooth bipolar shape, the later peak being smaller in amplitude to the earlier one. It was found that in some cases, particularly in the ones identified as having cervical spondylotic neuropathy based on DFL and MRI, there are deviations in the form of kinks in the waveshapes. The present work attempts to introduce a fair approximation of disorders by identifying those cases with kinks in their evoked EMG waveforms. To do this we categorize the collected signals into two groups:

Group A: having standard shape, which relates to normal healthy nerve.

Group B: having kinks in the evoked EMG M-response, which have been assumed to be related to neurological disorders, particularly of cervical spondylotic neuropathy.

The M-response (named as analysed signals in the figures) were available in discrete time sequence form; consisting of 512 points with sampling intervals of 40 μs.

Initially the first differentiations of the signals were studied, which, as mentioned earlier, has significant contributions from the rising and falling segments of the normal smooth curve, and from noise. Then the signals were analysed using CWT using MATLAB with its Wavelet Toolbox.

This paper concentrates on a specific category of continuous wavelets, Haar wavelets, which is just the simple step function satisfying the necessary conditions for Wavelet Transformation (WT).

Time-scale diagram and variation in the coefficient ($C_{a,b}$) line in CWT of the signals were studied. In the translation-scale diagram the vertical axis represents the dilation (scaling factor) of the wavelet, and the horizontal axis, its translation (shift along the time axis). Thus the wavelet transform plot can be seen as a color pattern against a two dimensional plane. In the WT plane each variation - the peak, valley and kinks on the evoked EMG responses is displayed as sharp dark lines, so presence of each sharp dark lines were located. The coefficient lines at midpoint of the range of scales, were also studied which relates to the depth of the color.

Evoked EMG signals were obtained for stimulation at two points on the Median nerve, typically at the wrist (Distal) and at the elbow (Proximal) giving two conduction distances. The dark lines of the WT were separately evaluated for responses obtained from both distal and proximal stimulations for both Group A and Group B.

A total of 36 evoked EMG M-responses from median nerves, for both distal and proximal stimulations, from 6 normal subjects demonstrating smooth shapes, and from 12 subjects demonstrating kinks in the evoked EMG wave form were studied. The latter were assumed to have neurological disorders as mentioned before.

## 3. RESULTS AND OBSERVATIONS

The output of the first order differentiations of a typical evoked EMG signal with kinks is shown in Fig.1. It can be seen that large values appears due to the rising and falling segments of the main curve while that due to the kinks in the middle are smaller. Besides, there are some noise spikes as well. Therefore, the kinks cannot be identified clearly from this first differentiation pattern

The CWT with Haar wavelets of evoked M-responses of two typical samples each from Group A (subject A1, distal and proximal stimulation) and Group B (subject B1, distal and proximal stimulation) are shown in Figs.2-5.

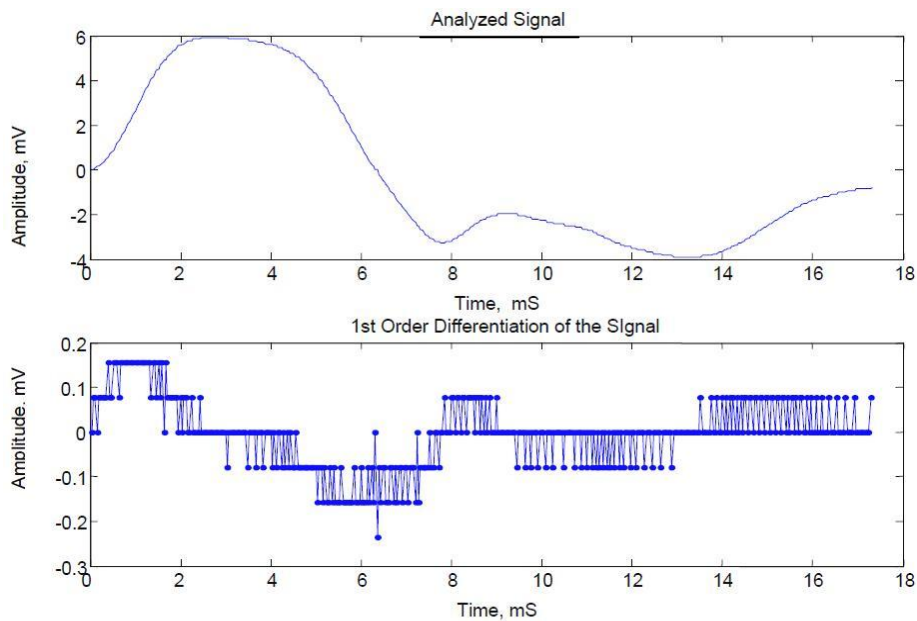

Figure 1: A CMAP signal and its corresponding first differentiation

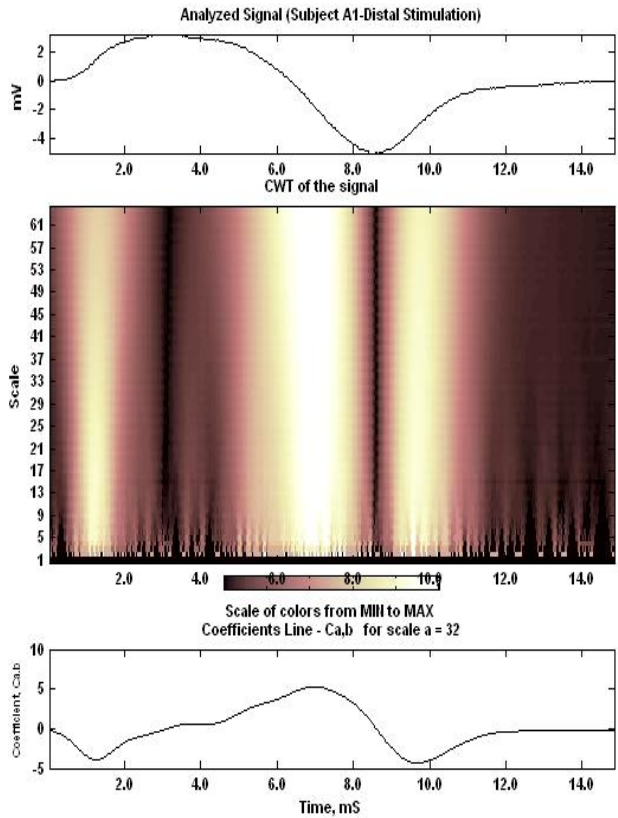

Figure 2: CWT with coefficient line variations of CMAP signals from a normal subject (subject A1 in Group A); for distal stimulation. Top curve: M-response (analysed signal), middle picture:: CWT of M-response, bottom curve: co-efficient line variation at the middle of the vertical axis.

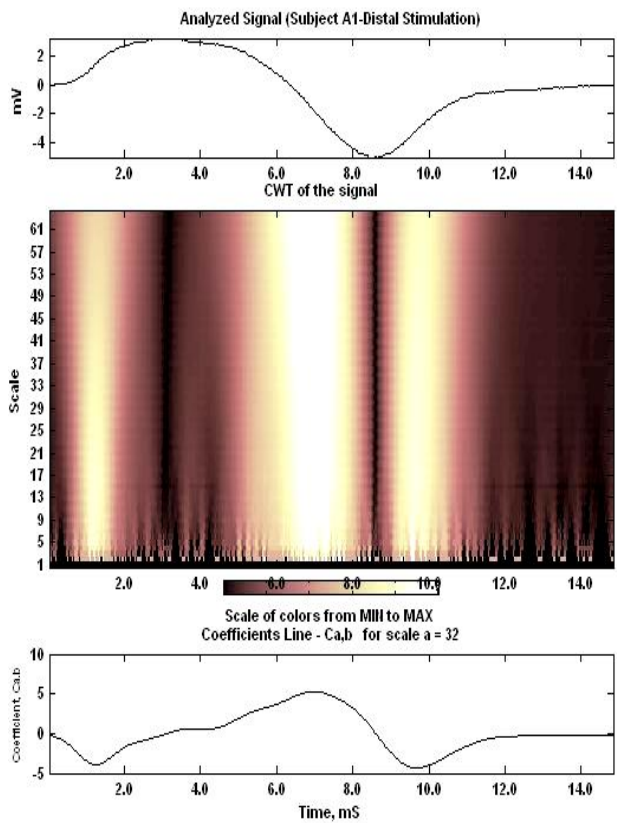

Figure 3: CWT with coefficient line variations of CMAP signals from a normal subject (subject A1 in Group A); for proximal stimulation. Top curve: M-response (analysed signal), middle picture:: CWT of M-response, bottom curve: co-efficient line variation at the middle of the vertical axis.

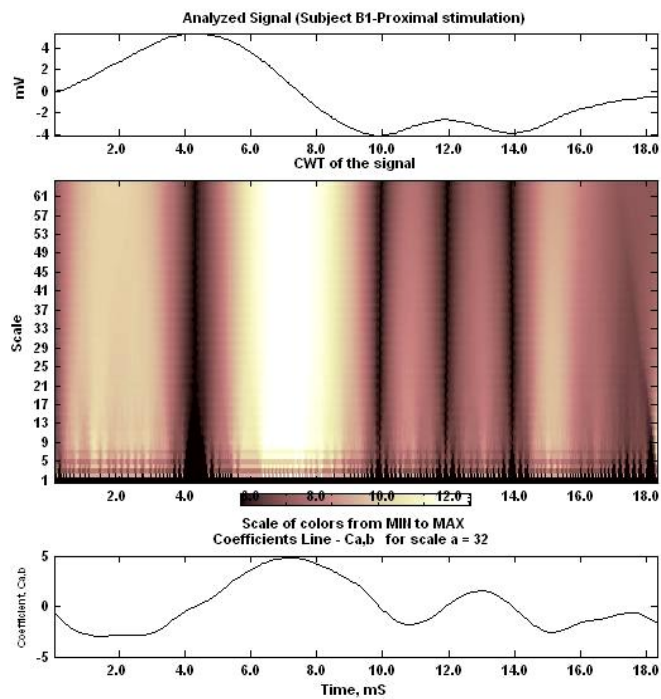

Figure 4: CWT with coefficient line variations of CMAP signals from an patient (subject B1 in Group B); for distal stimulation. Top curve: M-response (analysed signal), middle picture:: CWT of M-response, bottom curve: co-efficient line variation at the middle of the vertical axis.

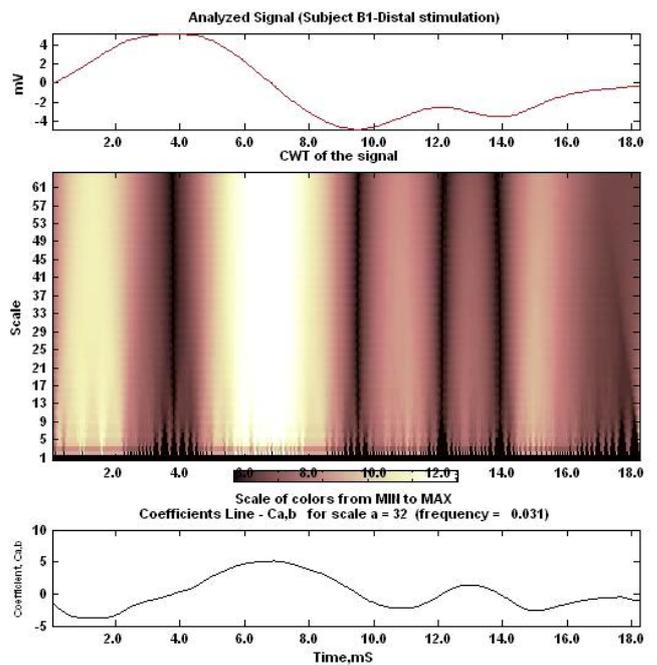

Figure 5: CWT with coefficient line variations of CMAP signals from a patient (subject B1 in Group B); for proximal stimulation. Top curve: M-response (analysed signal), middle picture:: CWT of M-response, bottom curve: co-efficient line variation at the middle of the vertical axis.

Plots show the evoked M-responses and the corresponding wavelet transform. The kinks of the waveform clearly correspond to dark line structures on the WT display unaffected by any noise in the original signal. This figure highlights the wavelet transform's ability to identify short duration, high frequency components in the time-frequency plane. When the scale factor, *a*, is enlarged, the effect on frequency is compression, i.e. the analysis window in the frequency domain is contracted by an amount of $1/a$ [28]. This equal and opposite frequency domain scaling effect is advantageous for frequency localization. Table 1 presents the values of locations of dark lines in the CWT plot for distal stimulation while Table 2 presents the same for proximal stimulation for the same subjects.

Table 1- Presence of the dark lines in CWT display
(distal stimulation)

| Subjects (Group A) | Dark Lines (ms) |
|---|---|
| A1 | 3.09, 8.59 |
| A2 | 2.90, 8.17 |
| A3 | 3.03, 8.23 |
| A4 | 4.35, 8.20 |
| A5 | 2.49, 7.28 |
| A6 | 1.54, 5.75 |
| Subjects (Group B) | Dark Lines (ms) |
| B1 | 2.27, 7.89, 8.64, 11.0 |
| B2 | 0.49, 2.71, 10.8 |
| B3 | 3.94, 9.64, 12.2, 14.0 |
| B4 | 3.75, 9.50, 12.1, 13.9 |
| B5 | 4.61, 5.02, 5.55, 12.3 |
| B6 | 2.85, 7.82, 9.22, 13.2 |
| B7 | 2.78, 7.77, 9.33, 13.3 |
| B8 | 1.94, 4.47, 5.94, 7.52, 8.08, 11.3 |
| B9 | 2.68, 10.0, 13.2, 15.0 |
| B10 | 2.08, 2.98, 9.33, 11.7, 13.0 |

Table 2- Presence of the dark lines in CWT display
(proximal stimulation)

| Subjects (Group A) | Dark Lines (ms) |
|---|---|
| A1 | 2.94, 8.51 |
| A2 | 3.17, 8.47 |
| A3 | 4.48, 8.37 |
| A4 | 4.35, 8.20 |
| A5 | 2.15, 7.30 |
| A6 | 1.87, 5.89 |
| Subjects (Group B) | Dark Lines (ms) |
| B1 | 2.86, 8.27, 9.66, 11.7 |
| B2 | 5.47, 11.6, 14.6, 16.0, 19.8, 21.7 |
| B3 | 4.29, 9.95, 12.0, 14.0 |
| B4 | 4.32, 9.82, 11.9, 13.9 |
| B5 | 3.02, 6.18, 12.4, 15.4, 20.6 |
| B6 | 3.94, 7.92, 9.19, 12.8 |
| B7 | 3.96, 7.70, 8.98, 12.7 |
| B8 | 3.22, 5.80, 7.28, 9.15, 10.1 |
| B9 | 4.0, 7.94, 9.27, 12.6 |
| B10 | 2.04, 3.29, 9.40, 11.5, 13.5, 17.4 |

For each of the subjects in Group A, CWT display has two common dark lines which correspond to the peak and valley of the normal evoked M-responses. It can be seen that those from Group B, having kinks in their evoked M-responses, have more than two dark lines at different time positions in WT displays as expected. More than two dark lines indicate the presence of kinks in the evoked M-response and would indicate neural disorders.

## 5. DISCUSSION

The present work is to provide an analysis of evoked EMG responses using Continuous Wavelet Transform, which have a strong capability to find the local variations in a signal, hence overcome the limitations of Fourier Transform.

Analyzed evoked M-responses served by the median nerve were obtained from the Thenar muscles in the palm. Normal evoked EMG signals usually have smooth bipolar shape and signals having kinks were found to be associated with neural disorders, and a preliminary work by the authors' group suggests that it may relate to cervical spondylosis, a very common disorder.

Kinks are high frequency components at different phases of the evoked M-responses, and thus, linear two dimensional time-scale displays were chosen for this study with an expectation that the dark lines (in contrast with the bright portion) may represent the positions of the kinks precisely. The Haar CWT techniques which are appropriate with bipolar shaped kinks were used as the first step towards wavelet transform. The coefficient lines were also studied to elucidate the color variations and to highlight the kinks more precisely.

The normal evoked M-responses have a common peak and a valley, thus two dark lines in the wavelet transformed display are common, presence of more than two dark lines ensures the existence of kinks, and hence confirms the neurological disorders. The Wavelet Toolbox of MATLAB was used to make the analysis easier.

Separation of the data in Group A and Group B depends on the shape of the signals. Selection of normal data under Group A had to be done under some constraints. Not many data sets were available which matched the perfect assumed shape.

The present work is noteworthy in that it has targeted development of a new approach to identify kinks in the evoked M-responses using computational techniques, so that diagnosis for neurological disorders may become simpler. Obviously Further work is necessary to establish the specific disorders that the kinks represent. However, the present work may be useful to isolate such cases. As stated before, there is a possibility of involving these kinks to cervical spondylosis (CS), a disorder leading to the disability of a great number of people globally. In that perspective the present work would be an important footstep for its diagnosis.